\documentclass[aps,prl,twocolumn,nofootinbib,groupedaddress,superscriptaddress
,longbibliography]{revtex4-2}
\AtBeginDocument{
\heavyrulewidth=.08em
\lightrulewidth=.05em
\cmidrulewidth=.03em
\belowrulesep=.65ex
\belowbottomsep=0pt
\aboverulesep=.4ex
\abovetopsep=0pt
\cmidrulesep=\doublerulesep
\cmidrulekern=.5em
\defaultaddspace=.5em
}

\usepackage{float}
\usepackage{comment}
\usepackage{booktabs}
\usepackage[percent]{overpic}
\usepackage{microtype}
\usepackage{array}
\usepackage{physics}
\usepackage{graphicx}
\usepackage{bbold}
\usepackage[usenames,dvipsnames]{xcolor}
\usepackage{tikz}
\usepackage{epsfig}
\usetikzlibrary{arrows.meta, automata, positioning, quotes, shapes}

\bibpunct{[}{]}{,}{n}{}{}

\definecolor{ForestGreen}{HTML}{668000}
\definecolor{red1}{HTML}{FF4136}
\definecolor{green1}{HTML}{00802b}

\usepackage{amsmath,amssymb}
\usepackage{multirow}
\usepackage{bm}
\usepackage{mathtools}
\usepackage{dsfont}
\usepackage{amsfonts}
\usepackage[normalem]{ulem}
\usepackage{url}
\usepackage{wasysym}
\usepackage{bbm}
 
\usepackage[colorlinks=true,linkcolor=blue,citecolor=blue,hypertexnames=false]{hyperref}

\def\ba#1\ea{\begin{align}#1\end{align}}
\def\bg#1\eg{\begin{gather}#1\end{gather}}
\def\bpm{\begin{pmatrix}}
	\def\epm{\end{pmatrix}}

\newcommand{\ourtitle}{Short-range Spin Freezing State in the Double Trillium Lattice Spin-Liquid Candidate KSrFe$_2$(PO$_4$)$_3$ Revealed via $^{31}$P NMR}
\allowdisplaybreaks

\begin{document}
\title{\textbf{\ourtitle}}
	
\author{Sebin J. Sebastian }
\affiliation{Ames National Laboratory, U.S. DOE, Iowa State University, Ames, IA 50011, USA}
\affiliation{Department of Physics and Astronomy, Iowa State University, Ames, IA 50011, USA}
\affiliation{School of Physics, Indian Institute of Science Education and Research Thiruvananthapuram-695551, India}
\author{Q. -P. Ding }
\affiliation{Ames National Laboratory, U.S. DOE, Iowa State University, Ames, IA 50011, USA}
\author{A. A. Tsirlin}
\affiliation{Felix Bloch Institute for Solid-State Physics, Leipzig University, 04103 Leipzig, Germany}
\author{R. Nath}
\email{rnath@iisertvm.ac.in}
\affiliation{School of Physics, Indian Institute of Science Education and Research Thiruvananthapuram-695551, India}
\author{Y. Furukawa}
\email{ furukawa@ameslab.gov}
\affiliation{Ames National Laboratory, U.S. DOE, Iowa State University, Ames, IA 50011, USA}
\affiliation{Department of Physics and Astronomy, Iowa State University, Ames, IA 50011, USA}

\date{\today}
\begin{abstract}
A comprehensive $^{31}$P nuclear magnetic resonance (NMR) study, combined with thermodynamic measurements and first-principle band-structure calculations, has been conducted to explore the ground state of the $S = 5/2$ double trillium lattice antiferromagnet KSrFe$_2$(PO$_4$)$_3$. 
Our experimental results indicate that the magnetic ground state is neither a conventional three-dimensional (3D) long-range order (LRO) nor a pure gapless spin-liquid state, as conjectured previously [Boya \textit{et al.}, APL Mater. \textbf{10}, 101103 (2022)]. Specifically, the observation of a nearly field-independent NMR linewidth below $T^{*} = (3.5 \pm 0.4)$~K, and a significant enhancement of spin-spin relaxation rate $1/T_2$ below $2T^{*}$ (where $T^{*}$ is the characteristic temperature identified from the magnetic susceptibility), indicate a complex magnetic ground state where spin freezing coexists with persistent dynamics. Furthermore, we argue that the lack of magnetic LRO and the persistence of strong magnetic fluctuations in KSrFe$_2$(PO$_4$)$_3$ are unlikely to originate from intersite K/Sr disorder, rather arise due to intrinsic magnetic frustration. Our findings position KSrFe$_2$(PO$_4$)$_3$ into a broader family of geometrically frustrated magnets characterized by coexisting spin freezing and pronounced antiferromagnetic fluctuations, marking it as a promising platform for investigating exotic phenomena in 3D frustrated magnets.
\end{abstract}

\maketitle
\let\oldaddcontentsline\addcontentsline
\renewcommand{\addcontentsline}[3]{}
Landau's proposition warrants that a phase transition in a many-body system should be accompanied by symmetry breaking, resulting in distinct phases at low temperatures~\cite{Sachdev33}. Interestingly, this convention is violated in the case of quantum spin liquid (QSL): a highly correlated but dynamically disordered state of spins characterized by fractionalized excitations, absence of magnetic long-range order (LRO) down to absolute zero temperature, and robust entanglement~\mbox{\cite{Balents199,Savary016502}}. While spin-1/2 two-dimensional (2D) triangular lattice antiferromagnets are subjected to extensive research following Anderson's prediction of the resonating-valence-bond (RVB) state known as a prototype QSL~\cite{Anderson153,li2020}, three-dimensional (3D) frustrated magnets also provide ample opportunities to harness QSL states. Up to now, only a handful number of 3D frustrated magnets have emerged as promising candidates for realizing QSL states; for example, hyperkagome lattices Na$_4$Ir$_3$O$_8$~\cite{Okamoto137207,Zhou197201,Lawler227201} and PbCuTe$_2$O$_6$~\cite{Khuntia107203,Hong256701,Chillal2348} and pyrochlore lattices NaCaNi$_2$F$_7$~\cite{Zhang167203,Plumb54}, LiGa$_{0.2}$In$_{0.8}$Cr$_4$O$_8$~\cite{Lee47}, and Ce$_2$Zr$_2$O$_7$~\cite{Gao1052}. 

Trillium lattice is another interesting addition to the field of complex 3D frustrated magnets. It comprises a 3D chiral network of three corner-sharing triangles~\cite{Hopkinson224441} and provides a promising framework for investigating frustration-induced quantum phenomena, including QSL~\cite{Ming2024,Fancelli2024}. The oxide-based trillium lattice compounds have garnered profound interest, especially after the recent discovery of a few systems manifesting fascinating quantum states. For instance, K$_2$Ni$_2$(SO$_4$)$_3$ (Ni$^{2+}$; $S = 1$) demonstrates a field-induced QSL driven by strong quantum fluctuations~\cite{Ivica157204,Yao146701,Gonzalez7191}. KBaCr$_2$(PO$_4$)$_3$ (Cr$^{3+}$; $S = 3/2$) exhibits a field-induced transition, while Na[Mn(HCOO)$_3$] ($S = 5/2$) is known for its unique 2-\textbf{k} magnetic ground state and a pseudo-plateau at $1/3$ of its saturation magnetization~\cite{Kolay224405,Bulled177201}. These examples highlight the diversity of magnetic states in trillium lattices, emphasizing the role of geometric frustration and hierarchy of magnetic interactions in shaping their ground states.

KSrFe$_2$(PO$_4$)$_3$ (KSFPO), a member of the same langbeinite family with the space group $P2_1 3$, presents a 3D double-trillium geometry of Fe$^{3+}$ ($S = 5/2$) ions. 
The dc magnetic susceptibility ($\chi_{\rm dc}$) follows the Curie-Weiss (CW) behavior with a large antiferromagnetic (AFM) CW temperature $\theta_{\rm CW} \simeq -70$~K and no sign of magnetic LRO down to 2~K. This absence of LRO is further corroborated by the absence of magnetic Bragg peaks in neutron diffraction measurements down to 2~K~\cite{Boya101103}. Additionally, the magnetic specific heat displays no evidence of LRO down to 0.19~K, follows a nearly quadratic temperature dependence ($\sim T^{2.3}$) at low temperatures, and remains field-independent up to 11~T. Based on these features, KSFPO was proposed to be a spin-liquid (SL) candidate on the double-trillium lattice~\cite{Boya101103}.

In this Letter, we present an in-depth investigation of the magnetic properties and spin dynamics of KSFPO using $^{31}$P nuclear magnetic resonance (NMR), dc and ac susceptibility, and specific heat measurements as well as density-functional theory (DFT) calculations. 
In contrast to the gapless SL state proposed earlier~\cite{Boya101103}, our experiments reveal a quasi-static frozen state coexisting with persistent AFM fluctuations in KSFPO, an emergent feature of 3D frustrated magnets, distinct from both spin-glass and conventional AFM order.

The polycrystalline sample of KSFPO used in this study was synthesized following the procedure outlined in Ref.~\cite{Hidouri145}, and phase purity was confirmed through high-resolution synchrotron x-ray diffraction. NMR measurements on $^{31}$P (\mbox{$I$ = $\frac{1}{2}$}, $\frac{\gamma_{\rm N}}{2\pi}$ = 17.2356 MHz/T) nuclei were conducted using a laboratory-built phase-coherent spin-echo pulse spectrometer. The details about the experimental protocols including NMR, dc and ac magnetic susceptibility, and specific heat measurements as well as sample preparation and DFT calculations are given in the Supplementary Material (SM)~\cite{SM}.

\begin{figure}
	\centering
	\includegraphics[width=\linewidth]{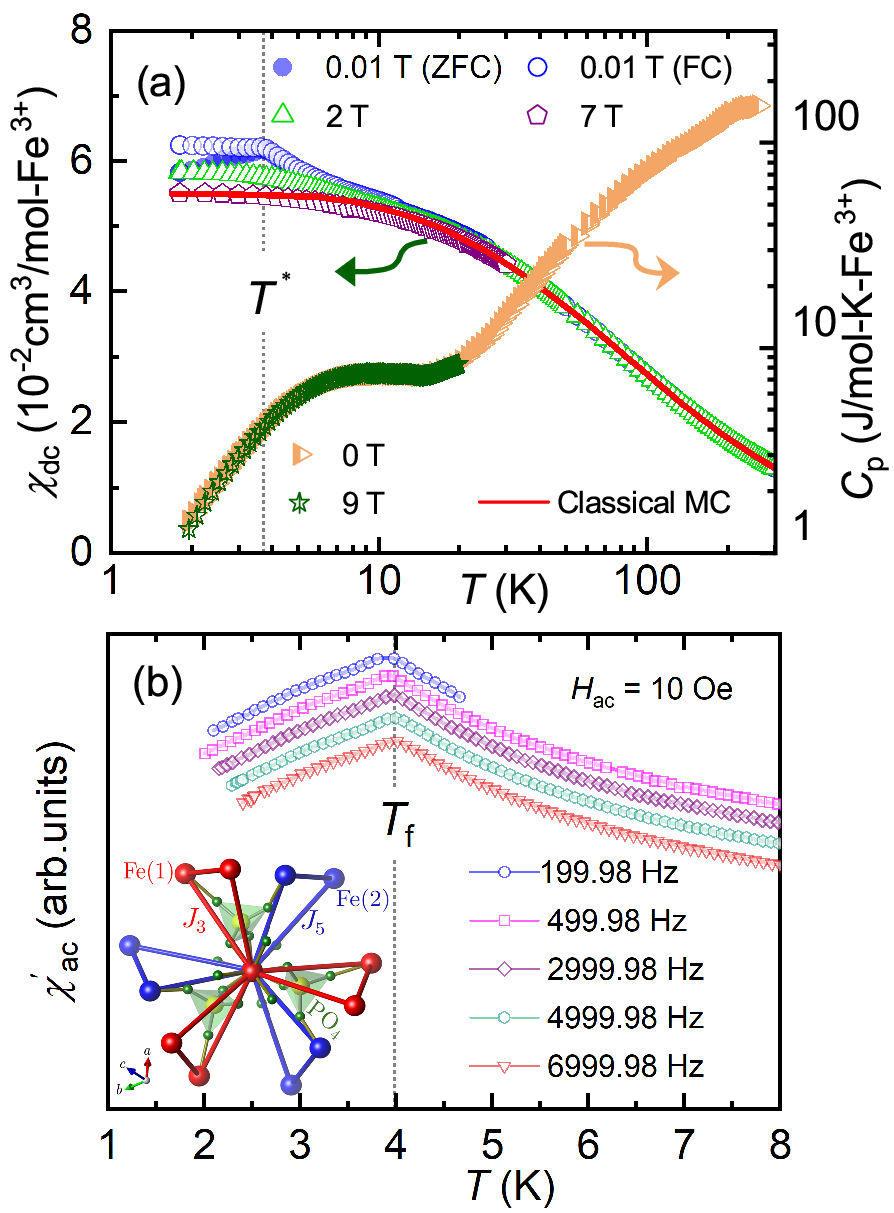}
	\caption{(a) Left $y$-axis: $\chi_{\rm dc}$ vs $T$ measured under ZFC and FC conditions in $\mu_0 H = 0.01$~T and in other applied fields. $T^*$ denotes the characteristic temperature. The solid line represents the classical Monte-Carlo simulation for the optimized exchange parameters from Table~\ref{tab:exchange}. Right $y$-axis: $C_{\rm p}$ vs $T$ measured in different magnetic fields. (b) Temperature-dependent $\chi^{\prime}_{\rm ac}$ measured at a fixed ac field of $H_{\rm ac}= 10$~Oe in different frequencies. The data at different frequencies are vertically offset for clarity. Inset: a partial view of the coupled trillium lattice of Fe$^{3+}$ ions, connected via PO$_4$ tetrahedra.}
	\label{Fig1}
\end{figure}
Figure~\ref{Fig1}(a) depicts the temperature-dependent $\chi_{\rm dc}$ measured in different applied magnetic fields. In a low magnetic field of 0.01~T, a weak bifurcation emerges between $\chi_{\rm dc}$ measured under zero-field-cooled (ZFC) and field-cooled (FC) conditions near the characteristic temperature, $T^{*} \simeq 3.5$~K which disappears in magnetic fields exceeding 1~T. However, no clear $\lambda$-type anomaly is observed in the specific heat [$C_{\rm p}(T)$] data [see Fig.~\ref{Fig1}(a)], which instead exhibits a very broad feature. The broad maximum in $C_{\rm p}(T)$ is also found to be field-independent, evocative of AFM short-range correlations, similar to the other 3D SL candidates, Na$_4$Ir$_3$O$_8$~\cite{Okamoto137207} and PbCuTe$_2$O$_6$~\cite{Koteswararao035141}. The real part of the ac susceptibility ($\chi^{\prime}_{\rm ac}$) exhibits a peak at around $T_{\rm f}\simeq 3.9$~K, slightly higher than $T^{*}$ determined from $\chi_{\rm dc}$ [see Fig.~\ref{Fig1}(b)]. This peak position is almost frequency-independent, which does not align well with typical spin-glass behavior~\cite{SM}. Additionally, magnetic isotherm $M(H)$ in the low-field regime recorded well below $T^{*}$ reveals no hysteresis, excluding the presence of ferromagnetic interactions, often required for spin-glass~\cite{SM}. Thus, the possibility of spin-glass formation in KSFPO is unlikely.


Further, it is observed that $\chi_{\rm dc}$ in the low-temperature regime evinces a tendency of saturation for $\mu_0 H \geq 1$~T. This feature does not support the presence of conventional AFM ordering at $T^{*}$. A Curie-Weiss (CW) fit to $\chi_{\rm dc}(T)$ at high temperatures ($T > 100$~K) returns an effective magnetic moment, $\mu_{\rm eff} = 5.96(2)\mu_{\rm B}$ (consistent with the spin-only value of $\mu_{\rm eff} \simeq 5.91\mu_{\rm B}$ for $S = 5/2$) and a CW temperature $\theta_{\rm CW} \simeq -70$~K. These findings are consistent with the results presented in the previous report~\cite{Boya101103}. Typically, for 3D frustrated and SL systems, the broad maximum in $C_{\rm p}(T)$ is expected to appear roughly in the range, $T/\theta_{\rm CW}\simeq 0.02-0.15$~\cite{Koteswararao035141}. Indeed, for KSFPO, this ratio is close to 0.065, which falls well within the expected range, indicating the proximity of KSFPO to the SL state.

\begin{table}
\caption{
\label{tab:exchange}
Exchange couplings $J_i$ (in K) in KSFPO reported in Ref.~\cite{Boya101103}, calculated in the present work using DFT, and optimized to achieve the best fit of the experimental magnetic susceptibility (Fig.~\ref{Fig1}). The labels follow Ref.~\cite{Gonzalez7191}. The last column shows the Curie-Weiss temperature $\theta_{\rm CW}$ (in K) averaged over the Fe1 and Fe2 sites.
}
\begin{ruledtabular}
\begin{tabular}{ccccccr}
                       & $J_1$ & $J_2$ & $J_3$ & $J_4$ & $J_5$ & $\theta_{\rm CW}$ \smallskip\\
Ref.~\cite{Boya101103} &  4.9  &  2.7  &  3.9  &  5.0  &  5.3  & $-162$ \\
DFT                    &  1.6  &  1.0  &  1.7  &  2.9  &  4.1  & $-89$ \\
Optimized              &  1.5  &  0.0  &  1.6  &  2.7  &  3.9  & $-76$ \\
\end{tabular}
\end{ruledtabular}
\end{table}

\begin{figure*}[h!tb]
\centering
\includegraphics[width=1.0\textwidth]{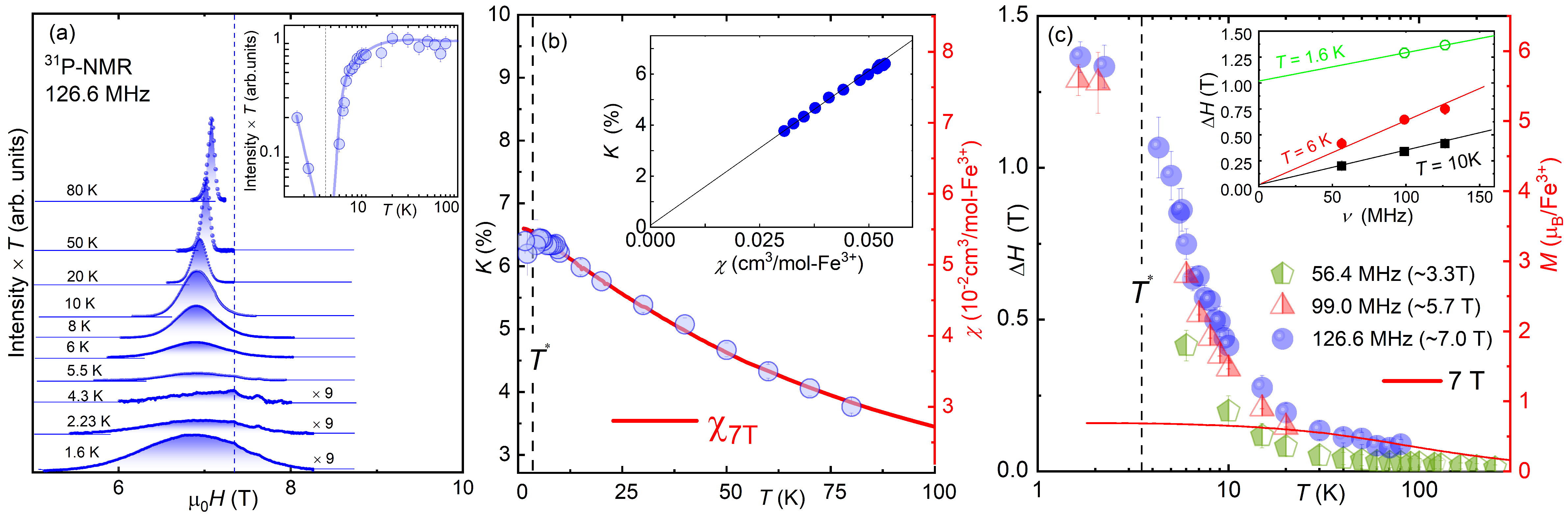}
\caption{(a) Temperature dependence of the field-swept  $^{31}$P NMR spectra measured at 126.6 MHz. The vertical dashed line marks the Larmor field. The inset shows the temperature dependence of the signal intensity multiplied by temperature with the $T_2$ corrections. (b) Temperature dependence of the $^{31}$P NMR shift ($K$) measured at 126.6 MHz, overlaid with the magnetic susceptibility data measured at 7 T. The inset shows $K$ vs $\chi_{\rm dc}$. (c) Full width at half maximum ($\Delta H$) of the $^{31}$P NMR line as a function of temperature, measured at three different frequencies. The solid line represents the magnetization ($M$) vs $T$ plot for $\mu_0 H = 7$~T. Inset: Variation of $\Delta H$ as a function of the NMR frequency, above and below $T^{*}$.}
\label{Fig2}
\end{figure*}

Our DFT calculations return five AFM exchange couplings (Table~\ref{tab:exchange}) that lead to a strong frustration of the double-trillium spin lattice. Not only the couplings within each trillium sublattice ($J_3$, $J_5$) are strong and AFM, in contrast to KBaCr$_2$(PO$_4)_3$~\cite{Kolay224405}, but also the couplings between the sublattices ($J_1$, $J_2$, and $J_4$) are sizable and enhance the frustration. While the AFM nature of all five couplings is in a qualitative agreement with the results of Ref.~\cite{Boya101103}, we note that the earlier values clearly overestimate the Curie-Weiss temperature of KSFPO. We performed classical Monte-Carlo simulations of the magnetic susceptibility and optimized the DFT-based exchange couplings using the experimental data (Fig.~\ref{Fig1})~\cite{vasp1,vasp2,pbe96,xiang2011,tsirlin2014,tsirlin2017,alps}. Excluding $J_2$ slightly improves the fit. By contrast, two other inter-sublattice couplings, $J_1$ and $J_4$, must be retained in order to describe the data. The dominant exchange couplings in KSFPO are $J_4$ and $J_5$ that should lead to the formation of a classical SL~\cite{Gonzalez7191}. However, sizable $J_1$ and $J_3$ are present as well.

We now employ the $^{31}$P NMR technique as a microscopic probe of the ground state. The temperature dependence of the $^{31}$P NMR spectra obtained by sweeping the external magnetic field ($H$) at a fixed frequency $\nu = 126.6$~MHz is shown in Fig.~\ref{Fig2}(a). Due to the unique crystallographic P-site in KSFPO, we observe a single NMR line which gradually broadens and moves towards the lower field side as the temperature decreases. As shown in the inset of Fig.~\ref{Fig2}(a), the signal intensity starts decreasing below 10 K and we were not able to observe the signal around  $T^{*}$ which is due to the extreme shortening of the nuclear spin-spin relaxation time $T_2$ beyond the limit of the NMR measurements as described below. Then, the signal intensity starts increasing at lower temperatures below $\sim 2$~K, but at 1.6~K it recovers only $\sim 20$\% of the expected intensity. Therefore, a large part of the signal is still missing at this temperature. This is very well evident from the intensity multiplied by temperature vs $T$ plot in Fig.~\ref{Fig2}(a), which is corrected by the $T_2$ effects. A similar wipe-out effect of the NMR signal intensity has been reported in other frustrated systems~\cite{Olariu167203}. It is noteworthy that the shape of the NMR line at the lowest temperature (below  $T^{*}$) remains Gaussian-like, which is quite different from a rectangular shape of the NMR spectrum expected in the LRO state of a conventional AFM. This suggests that the ground state of KSFPO is not a simple AFM LRO state.

Figure~\ref{Fig2}(b) shows the $T$ dependence of the NMR shift ($K$) determined from the peak position of the spectra as $K = [2\pi\nu/\gamma_{\rm N} - H_{\rm p}]/ H_{\rm p}$, where $H_{\rm p}$ represents the magnetic field of the peak position at each temperature. $K$ directly reflects the uniform magnetic susceptibility  $\chi_{\rm dc}$ as $K = K_0 + (\mathcal{A}_{\rm hf}/N_{\rm A})\chi_{\rm dc}$, where $K_0$ is the temperature-independent orbital contribution, $N_{\rm A}$ is the Avogadro's number, and $\mathcal{A}_{\rm hf}$ is the hyperfine coupling constant between the P nuclei and Fe$^{3+}$ spins. As shown in Fig.~\ref{Fig2}(b), $K(T)$ indeed tracks $\chi_{\rm dc}(T)$ measured at nearly the same magnetic field of 7~T very well. Such an agreement of $K$ with $\chi_{\rm dc}$ reflects the absence of any extrinsic contributions to the bulk magnetic susceptibility. Additionally, $K$ shows the tendency toward saturation at low temperatures, similar to $\chi_{\rm dc}$, thus clearly excluding a gapped ground state in this system. $\mathcal{A}_{\rm hf}\simeq 0.70$~T/$\mu_{\rm B}$ and $K_0 \simeq -0.13(1)\%$ were estimated from the linear fit to the $K$ vs $\chi_{\rm dc}$ plot with temperature as an implicit parameter, as shown in the inset of Fig. \ref{Fig2}(b).

It is worth noting that the full width at half maximum ($\Delta H$) exhibits a significant increase below 10~K, as shown in Fig.~\ref{Fig2}(c) where we plotted the $\Delta H$ measured at three different frequencies of 56.4, 99, and 126.6~MHz (corresponding to $\sim 3.2$, $\sim 5.74$, and $\sim 7$~T, respectively), together with the temperature dependence of magnetization $M$ measured at 7~T. As $\Delta H$ is proportional to the distribution of $\mathcal{A}_{\rm hf}M$,  $\Delta H$ is expected to be proportional to $M$ in the paramagnetic state for an NMR spectrum of a powder sample. In fact, a good scaling between $\Delta H$ and $M$ measured at $\sim 7$~T can be seen above 20~K. However, a substantial difference in the magnitude of $\Delta H$ and $M$ is observed below that temperature. As $\mathcal{A}_{\rm hf}$ does not change in this temperature range, which is guaranteed from the good scaling between $K$ and $\chi_{\rm dc}$, the large enhancement of $\Delta H$ is due to the local distribution of $M$ originating from the strong AFM correlations developing at low temperatures. Despite this large increase, $\Delta H$ is still proportional to $H$ above $T^*$, as expected in the paramagnetic state [see the inset of Fig.~\ref{Fig2}(c)]. On the other hand, below $T^*$, $\Delta H$ is found to be nearly independent of $H$, indicating a static internal field at the P site. By extrapolating the value of $\Delta H$ at 1.6 K below $T^{*}$ to zero NMR frequency ($\nu \to 0$), as shown in the inset of Fig.~\ref{Fig2}(c), we estimated the zero-field $\Delta H$ of $1.0 \pm 0.1$~T. This clearly indicates the freezing of Fe$^{3+}$ moments in the magnetic state below $T^*$. Given no LRO is detected in the specific heat measurements, our data suggest a quasi-static short-range-ordered state. The observation of the freezing of Fe$^{3+}$ moments below  $T^*$ excludes the formation of a pure dynamical SL state conjectured in the earlier study~\cite{Boya101103}.

\begin{figure}[tb]
\centering
\includegraphics[width=\linewidth]{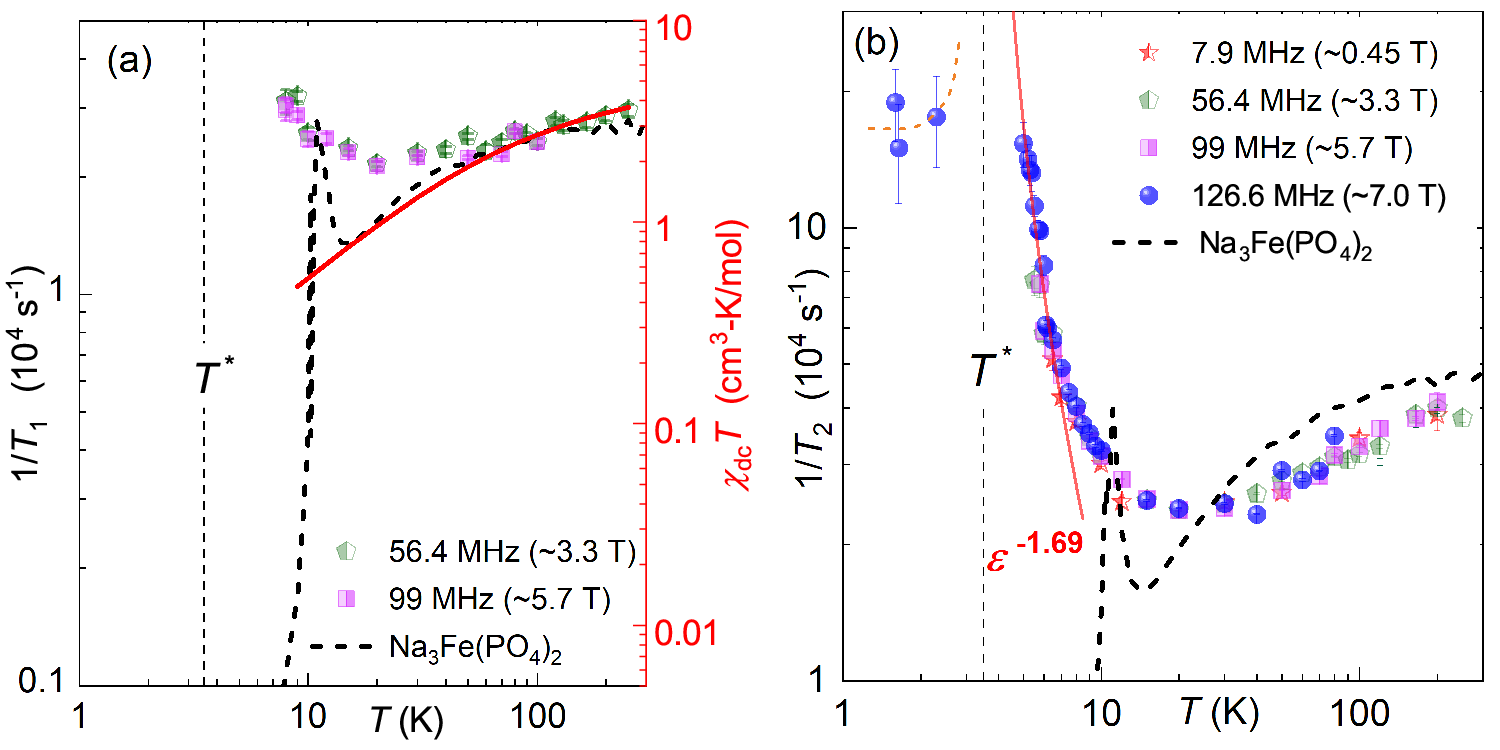}
\caption{(a) Temperature dependence of $1/T_1$ at different frequencies, together with the temperature dependence of $\chi_{\rm dc}T$ measured at 2 T. The black dashed curve represents the temperature dependence of $1/T_1$ reported for the typical conventional 3D AFM Na$_3$Fe(PO$_4$)$_2$ with a N\'eel temperature of $T_{\rm N}$ $\simeq$ 10.9 K from Ref.~\cite{Devi015803}. (b) The $^{31}$P $1/T_2$ as a function of temperature measured at various frequencies. The red solid line represents the fit discussed in the text, with $T^{*} = 3.5$ K. The orange dashed curve is included as a visual guide. The black dashed curve represents the temperature dependence of $1/T_2$ for Na$_3$Fe(PO$_4$)$_2$~\cite{Devi015803}.}
\label{Fig3}
\end{figure}

To obtain further detailed information about the spin dynamics, we measured the $^{31}$P spin-lattice relaxation rate ($1/T_1$) using the saturation-recovery method~\cite{SM}. The temperature-dependent $1/T_1$ are presented in Fig.~\ref{Fig3}(a). Here we measured $1/T_1$ at two frequencies: $\nu = 56.4$~MHz ($\sim 3.2$~T) and 99~MHz ($\sim 5.7$~T) and no obvious frequency dependence was found over the entire measured temperature range. With decreasing $T$, $1/T_1$ decreases and exhibits a local minimum around 20~K and then starts increasing. Below $\sim 8$~K, $T_1$ was too short to be measured.

In general, $1/T_1$ can be expressed in terms of the imaginary component of the dynamic susceptibility, $\chi\prime\prime (\vec{q},\omega_{\rm N})$ at the Larmor frequency $\omega_{\rm N}$ as~\cite{Moriya516,Nath214430} 
\begin{equation}
\dfrac{1}{T_1} = \dfrac{2\gamma^{2}_{\rm N}k_{\rm B}T}{N^{2}_{\rm A}} \sum_{\vec{q}}^{} |\mathcal{A}(\vec{q})|^2 \dfrac{\chi\prime\prime (\vec{q},\omega_{\rm N})}{\omega_{\rm N}}.
\label{eq1}
\end{equation}
Here, the sum is over the $\vec{q}$ vectors within the first Brillouin zone, and $\mathcal{A}(\vec{q})$ is the form factor of the hyperfine interactions. On the other hand, the uniform $\chi_{\rm dc}$ corresponds to the real component $\chi^{\prime}(\vec{q}, \omega_{\rm N})$ with $q = 0$ and $\omega_{\rm N} = 0$. Thus, the good scaling between $1/T_{1}$ and $\chi_{\rm dc}T$ at high temperatures above $\sim 50$~K shown in Fig.~\ref{Fig3}(a) clearly indicates that the $T$ dependence of $\sum_{\vec{q}}|\mathcal{A}(\vec{q})|^2\chi^{\prime\prime}(\vec{q}, \omega_{\rm N})$ scales to that of $\chi^{\prime}(0, 0)$.
This suggests that the nuclear relaxations can be explained by the simple paramagnetic fluctuations of Fe$^{3+}$ spins. Below $\sim$50 K, on the other hand, $1/T_1$ is greater than $\chi_{\rm dc}T$.  
This implies $\sum_{\vec{q}}|\mathcal{A}(\vec{q})|^2\chi^{\prime\prime}(\vec{q}, \omega_{\rm N})$ increases faster than $\chi^{\prime}$(0, 0), which is naturally attributed to a growth of spin fluctuations with $q$ $\neq$ 0, most likely with an AFM wave vector $\vec{q}$ = $\vec{Q}_{\rm AFM}$. On further cooling below $\sim$20 K, $1/T_1$ starts increasing. This could be due to the critical slowing down of spin fluctuations near $T^{*}$. However, the critical behavior of $1/T_1$ looks very different from the case of a conventional 3D AFM. We show this by plotting the $1/T_1$ data for the conventional antiferromagnet Na$_3$Fe(PO$_4$)$_2$ with $T_{\rm N} = 10.9$~K~\cite{Devi015803}, as an example. The critical behavior of $1/T_1$ is much sharper and is observed only close to $T_{\rm N}$ in Na$_3$Fe(PO$_4$)$_2$. In contrast, the critical behavior in KSFPO spans a relatively wide temperature range. Since the rapid shortening of $T_1$ makes $1/T_1$ measurements difficult at temperatures below 8~K, we measured the nuclear spin-spin relaxation time $T_2$ to further investigate the critical behavior close to $T^*$ and the spin dynamics in the low-temperature region.

The temperature dependence of $1/T_2$ measured at various frequencies is independent of frequency (magnetic field), as shown in Fig.~\ref{Fig3}(b). Similar to $1/T_1$, $1/T_2$ exhibits a weak temperature dependence and reaches a minimum around 20~K. Upon further cooling, $1/T_2$ shows a steeper increase below 10~K, again similar to $1/T_1$, which originates from the slowing down of the Fe$^{3+}$-spin fluctuations. It is worth noting that the magnitude of $1/T_2$ is very close to that of $1/T_1$ at low temperatures. This indicates that $1/T_2$ is determined by $1/T_1$. Therefore, one can examine in more detail the critical behavior near $T^*$ using the $1/T_2$ data. We analyzed the $T$ dependence of $1/T_2$ with  the formula
$1/T_2 \propto \epsilon^{-\beta} = A \left(\frac{T}{T^{*}}-1\right)^{-\beta}$, assuming $T^{*} = 3.5$~K. Here, $\beta$ represents the critical exponent and $A$ is a constant~\cite{Devi015803}. The resulting fit in the range of $T^{*}$ to $2T^{*}$ shown by the red curve in Fig.~\ref{Fig3}(b) yields a highly unusual value of the exponent $\beta\simeq1.69$, which is quite different from the reported values of $\beta = 0.38$ for the conventional AFM Na$_3$Fe(PO$_4$)$_2$~\cite{Devi015803} and $\beta = 0.5$ for the 2D unconventional triangular AFM NiGa$_2$S$_4$~\cite{Takeya054429}. For a comparison, we showed the typical temperature dependence of $1/T_2$ for Na$_3$Fe(PO$_4$)$_2$~\cite{Devi015803} by the black dashed curve in Fig.~\ref{Fig3}(b). It is clear that $1/T_2$ in KSFPO is strongly enhanced at low temperatures in comparison with the case of the 3D AFM Na$_3$Fe(PO$_4$)$_2$, although $1/T_2$ values are not much different at higher temperatures above $\sim 30$~K. Thus the strong enhancement of $1/T_2$ near $T^*$ and the large value of $\beta =1.69$ suggest the existence of intense and critical AFM fluctuations at low temperatures in KSFPO. These results exclude the formation of a conventional Heisenberg 3D AFM ordered state, consistent with the unusual short-range quasi-static ordered state revealed by the NMR spectral measurements.

It is interesting to point out that 1/$T_2$ is nearly constant below $T^*$. As described in the SM, while a single $T_2$ component was observed above $T^{*}$, a distribution of $T_2$ was detected below $T^{*}$, suggesting inhomogeneous magnetic fluctuations. The nearly $T$-independent behavior of $1/T_2$ together with quite large values of $1/T_2 \sim 1.5 \times 10^{5}$~s$^{-1}$ indicate the persistence of temperature-independent strong inhomogeneous magnetic fluctuations. This behavior contrasts sharply with a typical AFM order where $1/T_2$ or $1/T_1$ rapidly decreases with decreasing temperature inside the ordered state, as shown by the black dashed curves in Figs.~\ref{Fig3}(a,b).

The alkaline and alkaline-earth ions often show antisite disorder that can lead to structural randomness in the cubic langbeinite compounds. Our data suggest that this site disorder is weak in KSFPO~\cite{SM}. Moreover, the isostructural non-frustrated compound KBaCr$_2$(PO$_4$)$_3$ (Cr$^{3+}$; $S = 3/2$) with the much larger, about 30\% of the K/Ba intersite mixing undergoes an AFM LRO at 12.5~K in zero magnetic field followed by another transition at low temperatures in weak applied fields~\cite{Kolay224405}. This indicates that site disorder alone cannot explain the absence of AFM LRO in KSFPO, and we posit that magnetic frustration effects play a more significant role. Comparable quasi-static magnetic behavior has been identified in various other geometrically frustrated magnets. For instance, $\mu$SR and NMR studies of the hyperkagome Na$_4$Ir$_3$O$_8$ report a short-range spin-frozen state below $\sim~7$~K, coexisting with strong AFM fluctuations~\cite{Shockley047201,Dally247601}. Similarly, the Kitaev candidate Li$_2$RhO$_3$ exhibits a ZFC-FC bifurcation around 6~K and displays weak frequency dependence in its $\chi^{\prime}_{\rm ac}$, indicative of a quasi-static magnetic state~\cite{Khuntia094432}. Such observations reinforce that the coexistence of spin freezing and strong AFM fluctuations constitutes a distinctive hallmark of the ground states in these geometrically frustrated magnets.

In summary, our experimental study combined with first-principle calculations shows that KSFPO hosts an unconventional magnetic ground state characterized by short-range spin freezing and persistent strong AFM fluctuations. These findings stand in stark contrast to the earlier claim of a gapless SL ground state, although we confirm the presence of multiple competing AFM couplings that place KSFPO in the vicinity of the classical SL on the double-trillium lattice. We argue that the unusual behavior of KSFPO originates from the underlying magnetic frustration. KSFPO emerges as a particularly promising system for probing exotic magnetic ground states and thereby deepening our understanding of complex phenomena in 3D geometrically frustrated magnets.


\acknowledgments
We acknowledge D.V. Ambika for her valuable discussion and her assistance with the earlier stages of the NMR measurements. S.J.S acknowledges Fulbright-Nehru Doctoral Research Fellowship Award No.~2997/FNDR/2024-2025 and the Prime Minister's Research Fellowship (PMRF) scheme, Government of India to be a visiting research scholar at the Ames National Laboratory. S.J.S and R.N also acknowledges SERB, India, for financial support bearing sanction Grant No.~CRG/2022/000997 and DST-FIST with Grant No.~SR/FST/PS-II/2018/54(C). The research was supported by the U.S. Department of Energy, Office of Basic Energy Sciences, Division of Materials Sciences and Engineering. Ames National Laboratory is operated for the U.S. Department of Energy by Iowa State University under Contract No. DEAC02-07CH11358. We also thank ESRF for providing the beamtime for the synchrotron XRD measurement. 

\bibliography{References_KSFPO}
		
\end{document}